\documentstyle[osa,manuscript]{revtex}  

\newcommand{\ket}[1]{\left| #1 \right\rangle}
\newcommand{\mod}[2]{{#1}\hspace{0.7ex} {\rm mod}\hspace{0.7ex}{#2} }

\begin{document}                

\title{Concurrent Quantum Computation}

\author{F. Yamaguchi$^{\ast}$, C. P. Master$^{\ast}$ and Y. Yamamoto$^{\ast,\dagger}$}
\address{$^{\ast}$Quantum Entanglement Project, ICORP, JST\\
        Edward L. Ginzton Laboratory,
        Stanford University, Stanford, CA 94305, USA \\
        $^{\dagger}$NTT Basic Research Laboratories\\
        3-1 Morinosato-Wakamiya,
        Atsugi, Kanagawa 243-0198, JAPAN}
\date{}
\maketitle

\begin{abstract}                
A quantum computer is a multi-particle interferometer that
comprises beam splitters at both ends and arms, where the $n$
two-level particles undergo the interactions among them. The arms
are designed so that relevant functions required to produce a
computational result is stored in the phase shifts of the $2^n$
arms. They can be detected by interferometry that allows us to
utilize quantum parallelism. Quantum algorithms are accountable
for what interferometers to be constructed to compute particular
problems. A standard formalism for constructing the arms has been
developed by the extension of classical reversible gate arrays. By
its nature of sequential applications of logic operations, the
required number of gates increases exponentially as the problem
size grows. This may cause a crucial obstacle to perform a quantum
computation within a limited decoherence time. We propose a direct
and concurrent construction of the interferometer arms by one-time
evolution of a physical system with arbitrary multi-particle
interactions. It is inherently quantum mechanical and has no
classical analogue. Encoding the functions used in Shor's
algorithm for prime factoring, Grover's algorithm and
Deutsch-Jozsa algorithm requires only one-time evolution of such a
system regardless of the problem size $n$ as opposed to its
standard sequential counterpart that takes $O(n^3)$, $O(n)$ and
$O(n2^n)$.

\end{abstract}

\newpage

A computation entails encoding of a function whether classically
or quantum mechanically. The encoding of a function has been
carried out in the form of a bit-flip. Such an ``oracle" $U_{\rm
c}$  in reversible classical computers and also in proposed
quantum computers is a transformation of an $n$-bit input $x$ and
a work bit $w$,
\begin{equation}
    U_{\rm c}: \ket{x}\ket{w}\rightarrow\ket{x}\ket{w\oplus f(x)},
    \label{eq:U_c}
\end{equation}
where $\oplus$ stands for exclusive-OR. When the work bit is
initially set to be 0, the oracle returns the function value
$f(x)$ in the work bit. The customary construction of such an
oracle pertains to sequential application of reversible
gates\cite{Barenco95}. In quantum computation, however, the oracle
is often converted into a transformation of the form,
\begin{equation}
    U_{\rm q} :\ket{x}\rightarrow (-1)^{f(x)}\ket{x},
    \label{eq:U_q}
\end{equation}
where the information about $f(x)$ is encoded in the phase of a
linear superposition state so that quantum parallelism could be
utilized\cite{Deutsch85,Deutsch92,Grover97,Grover98}.
The conversion of $U_{\rm c}$ into $U_{\rm q}$ is performed by
supplementary transformation\cite{Deutsch85} or automatically by
appropriately initialized work bit\cite{Cleve98}. The construction
of the oracle (\ref{eq:U_c}) by sequential application of one-bit
and two-bit gates demand an exponential or polynomial number of
operations as the problem size grows as shown in
Table~\ref{table:scaling}. In order to perform a quantum
computation, we need to complete the construction of the oracle
while the coherence of the system is maintained, although
classical computation does not bring the issue of a limited
decoherence time into a question. We propose a concurrent
construction of the transformation $U_{\rm q}$ by only one-time
evolution of a physical system that has arbitrary multi-particle
interactions. The exponentially hard work that is expressed by the
complexity of gate arrays to be applied to a physical system in
the case of the sequential construction will be replaced by
adjustment of an exponentially large number of coupling strengths
in the system before the coherence of the system is required.

\newpage
The system required to concurrently implement an arbitrary $n$-bit
Boolean function $f(x)$ for an $n$-bit string $x$ consists of $n$
two-level particles with arbitrary multi-particle interactions
among them. The Hamiltonian of the system is,
\begin{eqnarray}
 {\cal H}&=&\sum_{i=1}^{n}\hbar\omega_i \sigma_{iz}
         +\sum_{i<j}\hbar\omega_{ij}\sigma_{iz}\otimes\sigma_{jz}
         +\sum_{i<j<k}\hbar\omega_{ijk} \sigma_{iz}\otimes\sigma_{jz}\otimes\sigma_{kz}\nonumber\\
         &&+\cdots +\hbar\omega_{12\cdots n}
         \sigma_{1z}\otimes\sigma_{2z}\otimes\cdots\otimes\sigma_{nz},
 \end{eqnarray}
where $\otimes$ stands for tensor product. The eigenstates of the
Pauli matrix $\sigma_{iz}=\left(\begin{array}{cc}
                    1 & 0 \\
                    0 & -1
\end{array}\right)$
for the $i$-th particle is used as the computational basis
$\{\ket{0}, \ket{1}\}$ for the $i$-th bit $x_i$, where
$\sigma_{iz}\ket{x_i}=(-1)^{x_i}\ket{x_i}$ ($x_i=\{0,1\}$). The
computational basis of $x$ is defined in the $2^n$-dimensional
state space spanned by the two-basis states of the $n$ particles,
$\ket{x}=\ket{x_{n}\cdots
x_1}=\ket{x_{n}}\oplus\cdots\oplus\ket{x_1}$. The state $\ket{x}$
represents a number $x=\sum_{i=1}^{n}2^{i-1}x_i$.
The Hamiltonian is diagonal in the computational basis
$\{\ket{00\cdots 0}, \ket{00\cdots 1},\cdots, \ket{11\cdots 1}\}$.
Diagonal elements of the $(2^{n}-1)$ terms in the Hamiltonian and
a $2^n$-dimensional vector $(1,1,\cdots,1)$ (diagonal elements of
the $2^n\times 2^n$ identity matrix) constitute an orthonormal
basis to expand $2^n$-dimensional vectors. It suggests that the
transformation $U_{\rm q}$ in (\ref{eq:U_q}), which is a
($2^n\times 2^n$)-diagonal matrix, can be constructed by a global
phase shift and one-time evolution of the system, $U_{\cal
H}(\tau) =e^{-\frac{i}{\hbar}{\cal H}\tau}$ for time $\tau$ as
\begin{equation}
    U_{\rm q}=e^{i\phi}U_{\cal H}(\tau).
    \label{eq:concurrent}
\end{equation}
The concurrent construction of the transformation $U_{\rm q}$ for
a given function comprises a preparation of the Hamiltonian (by
adjusting the coefficients, $\hbar\omega_i$, $\hbar\omega_{ij}$,
$\cdots$, $\hbar\omega_{12\cdots n}$) so that
(\ref{eq:concurrent}) is satisfied and a time-evolution by that
Hamiltonian. Only the latter process requires the conserved
quantum coherence of the system.

The diagonal elements of (\ref{eq:concurrent}) are decomposed as
\begin{eqnarray}
e^{i\pi f(x)} &=&
 e^{i\phi}\times\prod_{i}e^{-i\omega_{i}\tau(-1)^{x_i}}
 \times\prod_{i<j}e^{-i\omega_{ij}\tau(-1)^{x_i+x_j}} \nonumber\\
 &&\times\prod_{i<j<k}e^{-i\omega_{ijk}\tau(-1)^{x_i+x_j+x_k}}
 \nonumber\\
 &&\times\cdots\times e^{-i\omega_{12\cdots
 n}\tau(-1)^{x_1+x_2+\cdots x_n}},
 \label{eq:exp}
\end{eqnarray}
which gives the solution,
\begin{eqnarray}
 &&\phi =\frac{\pi}{2^n}\sum_{x}f(x),\nonumber\\
 &&\omega_{i}\tau = -\frac{\pi}{2^n}\sum_{x}(-1)^{x_i}f(x),\nonumber\\
 &&\omega_{ij}\tau =-\frac{\pi}{2^n}\sum_{x}(-1)^{x_i+x_j}f(x),\nonumber\\
 &&\qquad\qquad\vdots\nonumber\\
 &&\omega_{12\cdots n}\tau =
 -\frac{\pi}{2^n}\sum_{x}(-1)^{x_1+x_2+\cdots+
 x_n}f(x),
 \label{eq:Solution}
\end{eqnarray}
and determines the coefficients in the Hamiltonian to be prepared
so that only one-time evolution for time $\tau$ by the Hamiltonian
will construct $U_{\rm q}$ by itself for a given function $f(x)$.
Note that when $f(x)$ is a constant function, all coefficients in
(\ref{eq:Solution}) except $\phi$ are zero.

The formalism for concurrently constructing the transformation
$U_{\rm q}$ can be applied to encode functions and calculated
values that are used in existing quantum algorithms. Examples are
shown below.

\noindent {\it Deutsch-Jozsa
algorithm}\cite{Deutsch85,Deutsch92,Collins98}. The algorithm
solves the following problem by quantum parallelism. Given the
oracle for an $n$-bit Boolean function
$f:\{0,1\}^n\rightarrow\{0,1\}$, determine either (A) $f$ is a
constant function (at 0 or 1) or (B) $f$ is a balanced function
(the sequence $f(0)$, $\cdots$, $f(2^n-1)$ contains exactly
$2^{n-1}$ zeros and $2^{n-1}$ ones). In the original algorithm,
the oracle that has a form of $U_{\rm c}$ is used twice together
with another unitary operation on the work bit
$S:\ket{w}\rightarrow(-1)^{w}\ket{w}$ between the two
applications. The three unitary operations applied to the system
in series are designed so that the information about $f(x)$ is
transferred from the work bit to the phase of the control register
$\ket{x}$ as in (\ref{eq:U_q}). Now we have a method to construct
the transformation $U_{\rm q}$ concurrently, and the work bit can
be removed.

When $f(x)$ takes only 0 or 1, $f(x)$ in (\ref{eq:Solution}) can
be replaced by $-f(x)$ since $\pi f(x)\equiv -\pi f(x)$ (mod
$2\pi$). Therefore a set of parameters obtained by replacing
$f(x)$ by $(-1)^{x_1+x_2+\cdots+x_{n}}(f(x)-2 N_x)$ ($N_x$ is an
integer) in (\ref{eq:Solution}) is also solution to
(\ref{eq:exp}). For a balanced function, we choose $N_x=x_1 x_2$,
so that $\omega_{12\cdots n}\tau =
-\frac{\pi}{2^n}\left(2^{n-1}-\sum_{x}2N_x\right)=0$. Thus, to
encode a balanced $n$-bit Boolean function, we need the
multi-particle interactions up to $(n-1)$-th order (the
$n$-particle interaction is not required). Other parameters are
determined as,
\begin{eqnarray}
 &&\phi =\frac{\pi}{2^n}\sum_{x}(-1)^{x_1+x_2+\cdots+x_{n}}(f(x)-2 N_x),\nonumber\\
 &&\omega_{i}\tau = -\frac{\pi}{2^n}\sum_{x}(-1)^{x_1+\cdots+x_{i-1}+x_{i+1}\cdots+x_{n}}(f(x)-2 N_x),
    \nonumber\\
 &&\qquad\qquad\vdots.
\label{eq:Solution2}
\end{eqnarray}

\noindent {\it Grover's algorithm}\cite{Grover97,Grover98}. The
algorithm explains how a data can be found out of $2^n$ random
data entries. The data search problem is described by a function
$f(x)$ that returns 1 for a single unknown value of $x$, say
$x=\tau$,  and 0 for the rest of $x$. The algorithm uses the
information about $f(x)$ encoded in phase of the control register
as in (\ref{eq:U_q}), which can be implemented concurrently, when
the coefficients in the Hamiltonian are chosen as
$\phi=\frac{\pi}{2^n}$,
$\omega_i\tau=-\frac{\pi}{2^n}(-1)^{\tau_i}$,
$\omega_{ij}\tau=-\frac{\pi}{2^n}(-1)^{\tau_i+\tau_j}$,
$\cdots$,
and
$\omega_{12\cdots n}\tau
=-\frac{\pi}{2^n}(-1)^{\tau_1+\tau_2+\cdots\tau_n}$.

\noindent (iii) Simon's algorithm\cite{Simon97} determines whether
a given function $f:\{0,1\}^n\rightarrow\{0,1\}^m$ with $m\geq n$
is periodic $f(x)=f(x')\leftrightarrow x'=x\oplus s$ (with a
nontrivial string $s$) or one-to-one. Encoding the function $f(x)$
in the oracle of the form $U_{\rm c}$ in the original algorithm
can be replaced by
\begin{equation}
    \ket{x}\rightarrow e^{i\frac{\pi}{2^{m-1}}f(x)}\ket{x},
    \label{eq:Simon_U_q}
\end{equation}
which can be constructed concurrently. The transformation
(\ref{eq:Simon_U_q}) has the same form as $U_{\rm q}$, where
$(-1)=e^{i\pi}$ is replaced by $e^{i\pi/2^{m-1}}$. The formalism
for constructing the transformation $U_{\rm q}$ works exactly the
same way. In order to implement an arbitrary function in this
problem, all multi-particle interactions up to $n$-particle
interaction in the Hamiltonian are necessary.

\noindent {\it Shor's algorithm for prime
factorization}\cite{Shor97}. In order to factorize an odd number
$N$, we randomly choose $a$ ($N$ and $a$ need to be relatively
prime) and find the order $r$ of $a$, the least $r$ that satisfies
$a^r\equiv \mod{1}{N}$. Finding the order is the prime part of
Shor's algorithm and starts with the transformation on two $n$-bit
registers,
\begin{equation}
    \frac{1}{\sqrt{q}}\sum_{x=0}^{q-1}\ket{x}\ket{1}
    \rightarrow
    \frac{1}{\sqrt{q}}\sum_{x=0}^{q-1}\ket{x}\ket{\mod{a^x}{N}},
    \label{eq:Shor5.2}
\end{equation}
where $q=2^n$ satisfies $N^2\leq q <2N^2$. Replacing the
transformation (\ref{eq:Shor5.2}) by,
\begin{equation}
    \frac{1}{\sqrt{q}}\sum_{x=0}^{q-1}\ket{x}
    \rightarrow
    \frac{1}{\sqrt{q}}\sum_{x=0}^{q-1}e^{i\frac{\pi}{2N}(\mod{a^x}{N})}\ket{x},
    \label{eq:Shor5.2_mod}
\end{equation}
which encodes $\mod{a^x}{N}$ in the phase of the control register
instead of in the work bit, leaves the algorithm
unchanged\cite{modified-Shor}. The phase factors $\mod{a^x}{N}$ in
(\ref{eq:Shor5.2_mod}) can be calculated classically as,
\begin{equation}
    e^{i\frac{\pi}{2N}(\mod{a^x}{N})}
    = e^{i\frac{\pi}{2N}\prod_{i=1}^{n}(\mod{a^{2^{i-1}}}{N})^{x_{i}}},
    \label{eq:ShorU_q}
\end{equation}
where products refer to multiplication mod ($N$).
Equation (\ref{eq:ShorU_q}) has the same form as (\ref{eq:exp}) if
$\lambda_i^{x_i}=\frac{1+\lambda_i}{2}+\frac{1-\lambda_i}{2}(-1)^{x_i}$
($\lambda_i=\mod{a^{2^{i-1}}}{N}$) is used. Therefore, the
time-evolution of the system for time $\tau$ constructs the
transformation (\ref{eq:ShorU_q}) by itself when the coefficients
are chosen as,
$\phi=\prod_{i=1}^{n}\frac{1+\lambda_i}{2}$,
$-\omega_i\tau=\phi\times\frac{1-\lambda_i}{1+\lambda_i}$,
$-\omega_{ij}\tau=\phi\times\frac{1-\lambda_i}{1+\lambda_i}\times\frac{1-\lambda_j}{1+\lambda_j}$,
$\cdots$,
and $-\omega_{12\cdots
n}\tau=\prod_{i=1}^{n}\frac{1-\lambda_i}{2}$.

\noindent {\it Quantum Fourier transform.} To perform Quantum
Fourier transform $A_q$ for an $n$-bit register ($q=2^n$),
\begin{equation}
    A_q:\ket{x}\rightarrow\sum_{y=0}^{q-1}e^{2\pi ixy/q}\ket{y},
    \label{eq:QFT}
\end{equation}
as used in Shor's algorithm, we need the Walsh-Hadamard
transformation on each bit,
$
    H=\frac{1}{\sqrt{2}}\left(\begin{array}{cc}
        1 & 1\\
        1 & -1
    \end{array}\right)
$
in the computational basis, and controlled-phase-shift operators
on pairs of bits, defined as $S_{j,k}=e^{i\theta_{k-j}x_jx_k}$ on
the $j$-th bit and $k$-th bit with $j<k$ where
$\theta_{k-j}=\pi/2^{k-j}$. The right-hand side of (\ref{eq:exp})
is equal to $S_{j,k}$ if we choose
$2\omega_{jk}\tau=\theta_{k-j}$, $2\omega_{j}\tau=-\theta_{k-j}$,
$2\omega_{k}\tau=-\theta_{k-j}$ and the rest of the parameters are
zero, aside from the global phase factor. A product of multiple
controlled-phase-shift operators $S_{l,l'}S_{l,l'-1}\cdots
S_{l,l+1}$ is also diagonal in the computational basis and can be
implemented concurrently by setting
$2\omega_{l,m}\tau=\theta_{m-l}$ ($m=l+1,\cdots,l'$),
$2\omega_m\tau=-\theta_{m-l}$,
$2\omega_{l}\tau=-\sum_{m}\theta_{m-l}$. One-bit rotations and
two-particle interactions in the Hamiltonian are sufficient to
implement controlled-phase-shift operators and products of those
concurrently.

The time evolution of a system that concurrently implements
functions by our proposed scheme and the number of necessary
elementary gates for the function implementation by means of the
standard sequential scheme are compared in
Table~\ref{table:scaling} for existing quantum algorithm. It is
challenging to find a system that has multi-particle interactions
with reasonably large and controllable strengths in order to
implement arbitrary functions, but if found, many functions and
quantum algorithms will be implemented by only one-time evolution
of the system, which may cross out the current biggest obstacle to
quantum computation, the short decoherence time of a quantum
system.

\newpage

\begin{table}[t]
\newcommand{\lw}[1]{\smash{\lower2.0ex\hbox{#1}}}
\hspace*{-7ex}
\begin{tabular}{lccc}
    \hline
    & \multicolumn{2}{c}{Concurrent}
    & Sequential\\
    \lw{Algorithm/Function}& \multicolumn{2}{c}{implementation}
    & implementation \\
    \cline{2-4}
     & Terms in        & Evolution & Number of  \\
    & the Hamiltonian & time & elementary gates \\
    \hline
    & \multicolumn{3}{c}{$U_{\rm q}:\ket{x}\rightarrow(-1)^{f(x)}\ket{x}$}\\
    \hline
    General Boolean function $f(x)$ & {$RI^{(2)} \cdots I^{(n)}$} &  &
    \lw{$O(n2^n)$}\\
    Deutsch-Jozsa  & $RI^{(2)} \cdots I^{(n-1)}$ & $O(1)$ & \\
    Grover &  {$RI^{(2)} \cdots I^{(n)}$}&   & {$O(n)$} \\
    \hline
    & \multicolumn{2}{c}{$\ket{x}\rightarrow \exp\left[i\varphi{f(x)}\right]\ket{x}$}
    & {$\ket{x}\ket{0}\rightarrow \ket{x}\ket{f(x)}$}\\
    \hline
    Shor  ($\varphi=\frac{\pi}{2N}$, $f(x)=\mod{a^x}{N}$) & \lw{$RI^{(2)} \cdots I^{(n)}$} & \lw{$O(1)$} & $O(n^3)$
    \\
    Simon ($\varphi=\frac{\pi}{2^{m-1}}$) &  &   &  $O(mn 2^n)$\\
    \hline
    & \multicolumn{3}{c}{$S_{j,k}:\ket{x}\rightarrow
    \exp\left[i\frac{\pi}{2^{k-j}}x_kx_j\right]\ket{x}$}\\
    \hline
    controlled-phase-shift $S_{j,k}$&
    {$RI^{(2)}$} & {$O(1)$} & {$O(1)$} \\
    \hline
    \multicolumn{4}{r}{$H_{n}S_{n-1,n}H_{n-1}S_{n-2,n}S_{n-2,n-1}H_{n-2}\cdots
    H_{2}S_{1,n}S_{1,n-1}\cdots S_{1,3}S_{1,2}H_{1}$}
    \\\hline
    {Quantum Fourier Transform} & $RI^{(2)}$ & $O(n)$ & $O(n^2)$\\
    \hline
\end{tabular}
\caption{}\label{table:scaling}
\end{table}

\noindent Scaling laws of evolution time and number of elementary
gates necessary to implement functions for existing algorithms by
means of the proposed concurrent implementation and the standard
sequential implementation, respectively. In the most general case,
an $n$-bit Boolean function $f(x):\{0,1\}^{n}\rightarrow\{0,1\}$
is encoded in phases of the control register $\ket{x}$, as $U_{\rm
q}: \ket{x}\rightarrow(-1)^{f(x)}\ket{x}$, in order to utilize
quantum interference effect. To implement such an $n$-bit Boolean
function, we prepare the $n$-particle system that has
multi-particle interactions among them, from two-particle
interactions $I^{(2)}$ up to $n$-particle interactions $I^{(n)}$
in addition to one-bit rotations, $R$. A one-time evolution by the
Hamiltonian with properly chosen coefficients of the terms in it
builds the transformation $U_{\rm q}$ by itself. A Boolean
function $f(x)$ used in Deutsch-Jozsa problem\cite{Deutsch92} has
a constraint that $f(x)$ is either constant (at 0 or 1) or
balanced. Because of the constraint, the concurrent implementation
does not call for the $n$-particle interaction $I^{(n)}$. In
Grover's data search algorithm\cite{Grover97}, the function to be
implemented $f(x)$ is zero for all $x$ but $\tau$ ($f(\tau)=1$)
which we search for. As a general Boolean function, this function
requires the system that has all multi-particle interactions in
order to be implemented in a concurrent fashion. The
transformation $U_{\rm q}$ is constructed by successive
applications of elementary gates (one-bit gates and two-bit gates)
by the standard sequential implementation with the help of a work
bit prepared in the state $\frac{1}{\sqrt{2}}(\ket{0}-\ket{1})$
and another qubit\cite{Barenco95}. The number of required gate
operations to implement an $n$-bit Boolean function in the
sequential manner scales as $O(n2^n)$ as opposed to a one-time
evolution of a system in the case of the concurrent
implementation. The function implemented in Grover's algorithm is
special in that it only needs one $(n+1)$-bit gate that is
constructed by $O(n)$ elementary gates. In Shor's algorithm for
prime factoring an odd number $N$\cite{Shor97}, $\mod{a^x}{N}$
($a$ is a randomly chosen integer relatively prime to $N$) is
encoded in a work bit, as
$\ket{x}\ket{0}\rightarrow\ket{x}\ket{\mod{a^x}{N}}$. The
construction of a gate array for this transformation requires
$O(n^3)$ elementary gates by means of the sequential
implementation. Instead of encoding $\ket{\mod{a^x}{N}}$ is the
work bit, encoding it in the phase of the control register
$\ket{x}$, as
$\ket{x}\rightarrow\exp\left[i\frac{\pi}{2N}(\mod{a^x}{N})\right]\ket{x}$,
still works\cite{modified-Shor}. The factor $\frac{\pi}{2N}$ is
determined to differentiate all possible values $\ket{a^x}{N}$
takes (between 0 and $N-1$) and also to avoid a destructive
interference that ruins the algorithm. In Simon's
algorithm\cite{Simon97}, a function to be implemented is
$f:\{0,1\}^n\rightarrow\{0,1\}^m (m\geq n)$. The sequential
implementation of such a function in a work bit is in need of
$O(mn2^n)$ elementary gates. As in Shor's algorithm, the function
$f$ can be implemented in the phase of the control register
$\ket{x}$, as
$\ket{x}\rightarrow\exp\left[\frac{\pi}{2^{m-1}}f(x)\right]$. To
distinguish all $2^m$ possible values $f$ takes, the phase space
of $2\pi$ is divided by $2^m$ (a Boolean function,
$f:\{0,1\}^n\rightarrow\{0,1\}$, is a special case where $m=1$. In
both Shor's and Simon's algorithms, the concurrent implementation
of necessary functions involves all multi-particle interactions
$I^{(2)}$, $\cdots$, $I^{(n)}$ in the system. A
controlled-phase-shift operator $S_{j,k}$ acts on a pair of bits,
in this case $j$-th and $k$-th qubits of the control register
$\ket{x}$. It adds a phase factor
$\exp\left(i\frac{\pi}{2^{k-j}}\right)$ only when both $x_k$ and
$x_j$ are ones. Only two-particle interactions $I^{(2)}$ between
$k$-th and $j$-th particles to implement this operator. Such
controlled-phase-shift operators on all pairs of qubits
($n(n-1)/2$ paris in total) compose Quantum Fourier transform,
which is used in Shor's algorithm, along with the Walsh-Hadamard
transformation on each qubit ($H_l$ on $l$-th qubit)\cite{Shor97}.
The Walsh-Hadamard transformations are applied to all qubits, from
$x_1$ to $x_n$, and controlled-phase-shift operators
$S_{l,n}S_{l,n-1}\cdots S_{l,l+1}$ are interleaved between $H_l$
and $H_{l+1}$. A controlled-phase-shift operator and a product of
multiple of them can be concurrently implemented by a one-time
evolution of a system that has only two-particle interactions and
one-bit rotations. In total, Quantum Fourier transform is
constructed by $n$ Walsh-Hadamard transformations and $n$
time-evolutions of a system in a concurrent manner, whereas in the
case of the sequential implementation, $n(n-1)/2$
controlled-phase-shift operators are implemented in series.


\begin{thebibliography}{10}

\bibitem{Barenco95}
{A.~Barenco, C.~H.~Bennett, R.~Cleve, D.~P.~DiVincenzo,
N.~Margolus, P.~Shor,
  T.~Sleator, J.~A.~Smolin, H.~Weinfurter, Phys.~Rev.~A {\bf 52}, 3457 (1995).}

\bibitem{Deutsch85}
{D.~Deutsch, Proc.~R.~Soc.~London~A {\bf 400}, 97 (1985).}

\bibitem{Cleve98}
{R.~Cleve, A.~Ekert, C.~Macciavelo, M.~Mosca,
Proc.~R.~Soc.~London~A {\bf 454},
  339 (1998), quant-ph/970816.}

\bibitem{Deutsch92}
{D.~Deutsch, R.~Jozsa, Proc.~R.~Soc.~London~A {\bf 439}, 553
(1992).}

\bibitem{Collins98}
{D.~Collins, K.~W.~Kim, W.~C.~Holton, Phys.~Rev.~A {\bf 58}, R1633
(1998).}

\bibitem{Grover97}
{L.~K.~Grover, Phys.~Rev.~Lett. {\bf 79}, 325 (1997).}

\bibitem{Grover98}
{L.~K.~Grover, Phys.~Rev.~Lett. {\bf 80}, 4329 (1998).}

\bibitem{Simon97}
{D. R. Simon, SIAM J. Comput. {\bf 26}, 1474 (1997).}

\bibitem{Shor97}
{P. W. Shor, SIAM J. Comput. {\bf 26}, 1484 (1997).}

\bibitem{modified-Shor}
{Following Shor's original algorithm\cite{Shor97}, We then apply
the Quantum
  Fourier transform (\ref{eq:QFT}) to the control register $\ket{x}$ in the
  state (\ref{eq:Shor5.2_mod}), and then we obtain \begin{displaymath}
  \frac{1}{q}\sum_{x=0}^{q-1}\sum_{y=0}^{q-1}
  e^{i\frac{\pi}{2N}(\mod{a^x}{N})}e^{2\pi ixy/q}\ket{y}. \end{displaymath} A
  measurement projected to $\ket{y}$ has a probability, \begin{displaymath}
  \left|\frac{1}{q}\sum_{x=0}^{q-1} e^{i\frac{\pi}{2N} (\mod{a^x}{N})}e^{2\pi
  ixy/q}\right|^2. \end{displaymath} Since the order of $a$ is $r$,
  $\mod{a^x}{N}$ may be written as $a^k$, $0\leq k<r$, where $x\equiv
  \mod{k}{r}$. Therefore, the probability is, \begin{displaymath}
  \left|\sum_{k=0}^{r-1}
  e^{i\frac{\pi}{2N}(\mod{a^k}{N})}\frac{1}{q}\sum_{x:a^x\equiv a^k}e^{2\pi
  ixy/q}\right|^2 =\left|\sum_{k=0}^{r-1}
  e^{i\frac{\pi}{2N}(\mod{a^k}{N})}\frac{1}{q}\sum_{b=0}^{\lfloor
  (q-k-1)/r\rfloor} e^{2\pi i(br+k)yq}\right|^2. \end{displaymath} The sum
  taken over all $x$ satisfying $a^x\equiv \mod{a^k}{N}$, or equivalently
  $x\equiv \mod{k}{r}$, is replaced by the sum over $b$, defined by $x=br+k$,
  and it is decomposed into two terms. \begin{eqnarray*}
  \lefteqn{\frac{1}{q}\sum_{b=0}^{\lfloor (q-k-1)/r\rfloor} e^{2\pi
  i(br+k)y/q}}\\ &&=\frac{1}{q}\sum_{b=0}^{\lfloor (q-k-1)/r\rfloor} e^{2\pi
  ib\{ry\}_q/q} +\frac{1}{q}\sum_{b=0}^{\lfloor (q-k-1)/r\rfloor}\left[ e^{2\pi
  i(br+k)cq}-e^{2\pi ib\{ry\}_q/q}\right],\\ \end{eqnarray*} where $\{ry\}_q$
  is congruent to $\mod{ry}{q}$ ($-q/2 <\{ry\}_q \leq q/2$). The first sum is
  large only when $\{ry\}_q$ is small. When $|\{ry\}_q|\sim O(r)$, the second
  sum is $O(1/q)$ and the first term can be approximated as,
  \begin{displaymath} \frac{1}{r}\int_{0}^{1}\exp\left(2\pi
  i\frac{\{ry\}_q}{r}u\right) {\rm d}u + O\left(\frac{1}{q}\right),
  \end{displaymath} where the first term is $O(1/r)$. Neglecting terms $O(1/q)$
  ($\ll O(1/r)$), we obtain the probability of observing the state $\ket{y}$,
  \begin{displaymath} \left|\sum_{k=0}^{r-1}
  e^{i\frac{\pi}{2N}(\mod{a^k}{N})}\right|^2\times
  \left|\frac{1}{r}\int_{0}^{1}\exp\left(2\pi i\frac{\{ry\}_q}{r}u\right) {\rm
  d}u \right|^2. \end{displaymath} The second term is the probability of
  observing the state $\ket{y}$ for a particular value of $k$. If all possible
  values of $k$ ($0, 1, \cdots, r-1$) contribute the probability independently,
  the total probability is this term multiplied by $r$. In our case, the first
  term in the above equation is multiplied instead, due to the $k$-dependent
  phase factors. It can be calculated as, \begin{eqnarray*}
  \left|\sum_{k=0}^{r-1} e^{i\frac{\pi}{2N}(\mod{a^k}{N})}\right|^2
  &=&\sum_{k=0}^{r-1}\sum_{k'=0}^{r-1}e^{i\frac{\pi}{2N}(\mod{a^k}{N})}e^{-i\frac{\pi}{2N}a^{k'}}\\ &=& r +
  2\sum_{k=1}^{r-1}\sum_{k'=0}^{k-1}\cos\left[\frac{\pi}{2N}\left((\mod{a^k}{N%
})-(\mod{a^{k'}}{N})\right)\right]. \end{eqnarray*} Since
  $-N<(\mod{a^k}{N})-(\mod{a^{k'}}{N})<N$, the second term in the above
  equation is positive. The modified visibility $\left|\sum_{k=0}^{r-1}
  e^{i\frac{\pi}{2N}(\mod{a^k}{N})}\right|^2/r$ is lower-bounded by 1.}

\end{thebibliography}
\end{document}